\documentclass[11pt,twoside]{article}  
\usepackage{asp2006}
\usepackage{adassconf}

\begin{document}   

\paperID{10.2}
\title{Using Multipartite Graphs for Recommendation and Discovery}
\markboth{Kurtz et al.}{Using Multipartite Graphs}

\author{Michael J. Kurtz, Alberto Accomazzi, Edwin Henneken, Giovanni Di Milia,
Carolyn S. Grant}
\affil{Harvard-Smithsonian Center for Astrophysics, Cambridge, MA, USA}

\contact{Michael Kurtz}
\email{kurtz@cfa.harvard.edu}

\paindex{Kurtz, M. J.}
\aindex{Accomazzi, A.}
\aindex{Henneken, E.}
\aindex{Di Milia, G.}
\aindex{grant, C. S.}

\keywords{astronomy literature, network analysis, recommender sustems, eigenvector techniques}

\setcounter{footnote}{1}

\begin{abstract}          
The Smithsonian/NASA Astrophysics Data System exists at the nexus of a
dense system of interacting and interlinked information networks. The
syntactic and the semantic content of this multipartite graph
structure can be combined to provide very specific research
recommendations to the scientist/user.
\end{abstract}

\section{Introduction}

Modern information systems, such as the ADS (Kurtz, et al. 1993), act
as a nexus, linking together many densely interconnected systems of
information.  These systems can be viewed as systems of
interconnecting graphs, an example of a bipartite graph would be the
interaction of the set of all papers with the set of all authors,
which yields connections between papers and papers (papers are
connected if they have the same author) and between authors and
authors (co-authorship).

Modern computational techniques permit these rich data sources to be
used to solve practical problems.  Some techniques use the graph
representation to achieve orderings, such as with the Girvan-Newmann
algorithm (Girvan and Newmann, 2002) or the Rosvall-Bergstrom
algorithm (Rosvall and Bergstrom, 2008).  Others use eigenvector
techniques on the interconnectivity or influence matrices, either
using exact methods (e.g. Thurstone 1934, Ossorio 1965, Kurtz 1993),
or approximate methods suitable for huge systems such as PageRank
(Brin and Page 1998).

\section{A Faceted Browse System}

Developing practical solutions to the problem ``given my current state
of knowledge, and what I am currently trying to do, what would be the
best things for me to read'' requires an in depth understanding of the
properties of the data and the nature of the many different reduction
techniques.  The data are quite complex; as an example two papers (A
and B) can be connected to each other because 1) A cites B; 2) B cites
A; 3) A and B cite C; 4) Author X wrote both A and B; 5) Author X
wrote a set of papers, at least one of which was cited by both A and
B; 6) A and B were read by the same person; 7) A and B have the same
key word; 8) A and B refer to the same astronomical object; 9) etc..

A practical example of combining data and techniques would be to build
a faceted browse system for current awareness.  A possible avenue is:
take a set of qualified readers, say persons who read between 80 and
300 papers from the main astronomy journals within the last six
months; for each reader find the papers that reader read; for each of
these papers find the papers that that paper references; for each of
these papers find the keywords assigned to that paper by the journal;
next, for each reader create an N-dimensional normalized interest
vector, where each dimension is a keyword and the amplitude represents
the normalized frequency of occurrence in the papers cited by the
papers read.  This yields a reader-keyword matrix; one way to view
this is that the readers are points in a multidimensional keywork
space.

Several things can be done with this matrix, for example if the
readers are clustered, by K-means or some other algorithm, one obtains
groups of readers with similar interests.  These can be used as the
basis of a collaborative filter, to find important recent literature
of interest, and can be subdivided, to narrow the subject (as defined
by people with similar interests).  This creates a faceted browse of
important recent papers in subjects of current interest.  The ADS has
sufficient nubers of users to support three levels of facets.

\section{A Recommender System}

Using similar techniques the ADS is currently implementing a
recommender system: given that you are reading a particular article,
what other articles might be useful for you to read?  The first step
in developing a recommender system is to find a set of papers similar
to the paper being read.  A group of similar articles substantially
enhances the signal to noise of the recommender system compared with a
single article; the mean downloads per month for a ten year old
Astrophysical Journal article, for example, is one.  There are three
basic steps to this process: first create a system to find similar
articles for an arbitrary article; then given an article of interest
find those similar articles, and finally use them to find recommended
articles.  The system must be designed so that the recommended
articles can be chosen in real time, once the arbitrary article of
interest is selected.

One effective method for finding similar articles is: 1) take the
reader-keyword matrix and reduce its dimensionality (to about 50)
using SVD; 2) transform all the papers into the reduced dimensionality
system by fitting their keyword vectors to the significant SVD
vectors; 3) cluster the (50 dimensional) article keyword vectors into
many clusters (of about 1000 articles each) using hierarchal
clustering techniques; 4) for each of the small clusters of papers
perform a new SVD decomposition on the $\approx$50 dimensional
vectors, reducing the dimensionality further (to about 5); 5) for each
small cluster of papers transform each paper into the corresponding 5
dimensional subspace.  These steps can be done in advance as part of
the indexing necessary for a text retrieval system.

Now, to find suggested reading for a new article, say one just
released on arXiv, is relatively simple.  First the keyword vector
must be created for the paper, but note that this is a function of the
articles the paper references, the arXiv paper itself need not be
keyworded.  Next perfprm the transformations and classifications to
put the article into the proper small cluster of papers with its five
dimensional subspace.  Then find those n papers (40 is a reasonable
number) from the 1000 or so in the cluster which are closest in the
5-dimensional space to the input arXiv article.  These 40 articles
become the basis for the recommender systems.

We look for recommended papers using second order operators (Kurtz
1992, Kurtz, et al 2005).  For three possible recommendations we use
the betweenness of the papers in the group of 40; we find the paper
which was most read immediatly following the reading of a member of
the group, the paper most often read immediatly before a member of the
group, and the paper most often read either before or after (these can
be, and often are, different).  This inverts the concept of
betweenness centrality, we do not find the papers which are most
between a set of papers, we find the set of papers for which the group
of 40 papers very similar to our input paper are between.

Two additional recommendations can be gotten by finding all the people
who read any of the group of 40 papers, and finding the paper they
read the most in the last few months, and by finding the most recent
paper in the top 100 of this most alsoread list.  The citations can
give two more recommendations: the paper which the group of 40 papers
cite the most, and the paper which cites the largest number of the
group of 40.

Finally a joint query of ADS with SIMBAD (Wenger, et al. 2000) can
find the paper which refers to the largest number of astronomical
objects which are referred to by the papers in the group of 40 very
similar papers to the input paper.

\begin{figure}[t]
\epsscale{0.90}
\plotone{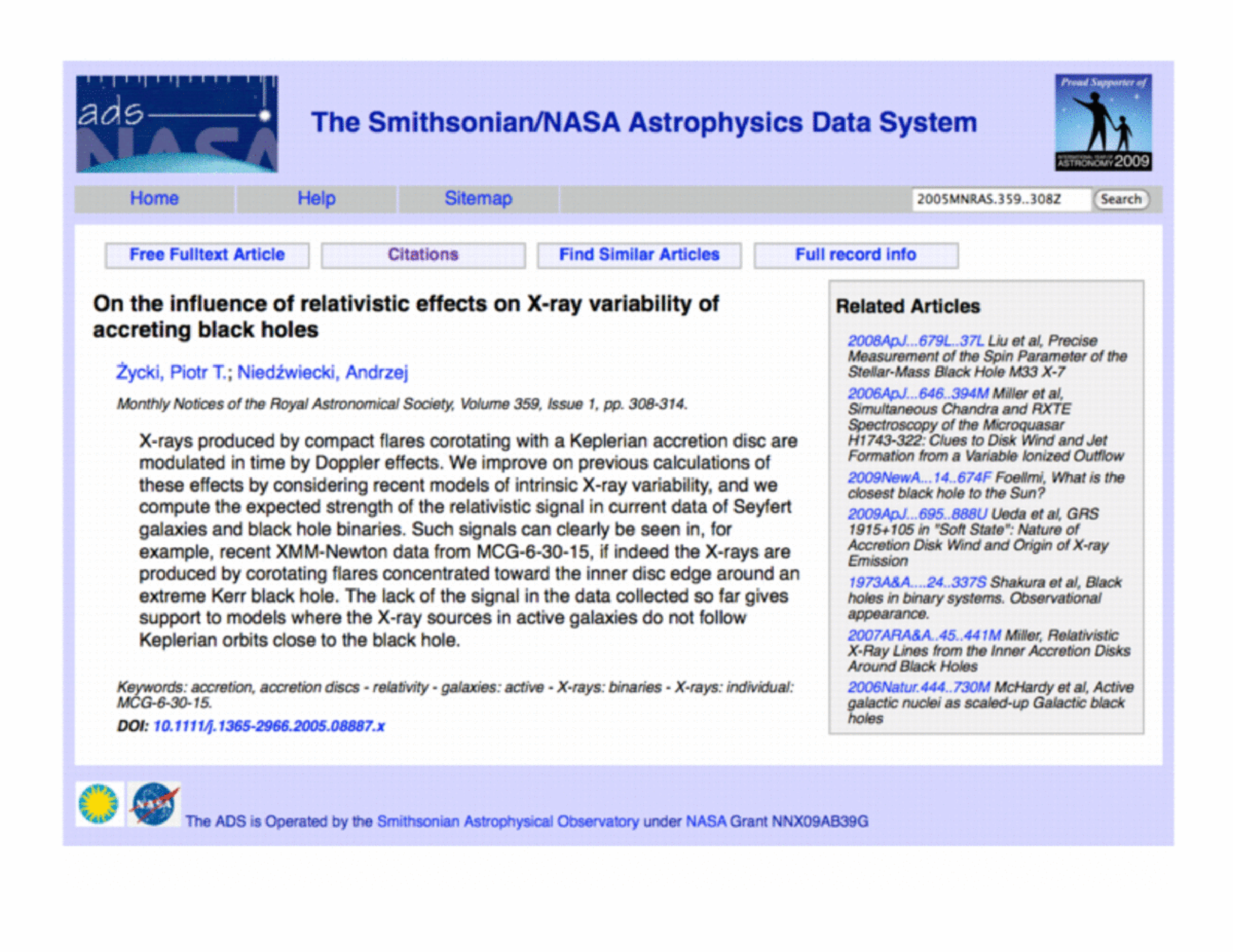}
\caption{An abstract with recommended articles.} \label{O4.1-fig-1}
\end{figure}

\section{Conclusion}

Clearly this is not the only way to find recommended papers; like with
architecture or civil engineering (there is no best building or bridge
design) the problem is too complex to be fully optimized.  There are
very likely ways of doing this which are better than others, however,
and this will be learned over time.

We only used part of the available data here; from the citations we
only used in-degree and out-degree from the group of 40.  We did not
use the author-based relations at all.  Instead of clustering the
papers based on a hierarchal clustering of the reduced keyword vector
(a subject matter technique) we could have clustered them based on the
co-ctation netword, using the Rosvall-Bergstrom algorithm (Kurtz, et
al. 2007).  We did not use any knowledge about the actual user. Etc.

These methosd are not restricted to scientific papers.  We are
entering an age of very densely interconnected ``data'' objects;
building intelligent systems to guide users in their traversal of
these new universes is clearly a branch of knowledge engineering whose
time has come.

\acknowledgments

The ADS is supported by NASA grant NCC5-189.


\end{document}